# Monte Carlo Modelling of Compton Scattering applied to Optimize Experimental Parameters

Assunta Sophia Felice, Imperial College London

By considering the effect of varying the target radii and detector aperture width on the scattering angle in experimental Compton scattering, mathematical models were developed and subsequently incorporated into Monte Carlo simulations. By simultaneously varying both of the investigated parameters, their optimal values were determined such that a minimum cumulative experimental uncertainty would be produced, accounting for the width of spectrographic photopeak and the number of detected counts; this minimum value was concluded to be 2.35 keV. For a fixed experimental set-up, including a fixed path length, these optimal values were concluded to be $0.0215 \pm 7 \times 10^{-4}$ m and $5.35 \times 10^{-3} \pm 2.34 \times 10^{-5}$ m for target diameter and aperture width, respectively. Recommendations for further adaptations to the investigation, such as incorporating attenuation coefficients and generalizing the simulation to three-dimensions are discussed.

## 1 Introduction

The Compton effect can be defined as the relativistic, inelastic scattering by light of a free charged particle. The wavelength of scattered light is not equal to that of the incident radiation, but is described instead by the Compton shift. This effect demonstrates that light cannot exist purely as a waveform; the classical theory of electromagnetic waves scattered by charged particles, Thomson Scattering, breaks down when considering shifts in wavelengths at low intensities. This effect should become arbitrarily small at low light intensities, regardless of wavelength, thus to fully explain low intensity Compton scattering, light must consist instead of particles [1]. This establishes that interactions between electromagnetic radiation and matter occur absorption or emission of discrete quanta. Experimental verification of momentum conservation in individual Compton scattering processes [2] was used to disprove Bohr-Kramers-Slater (BKS) theory, the interaction of matter and EM radiation using the classical wave description of quantum fields. Furthermore, it has practical applications in medical Physics: Compton scattering becomes the dominant process when human tissues are irradiated in the diagnostic and therapeutic radiation range, which is between 30keV and 30MeV.

The effect was first demonstrated upon the development of the technique of X-ray spectrometry to measure the wavelengths of scattered X-rays. The current experimental technique, as considered throughout this investigation, uses gamma ray spectrometry, as discussed in 2.2. The experimental technique was simulated computationally by use of Monte Carlo techniques; these were subsequently developed by considering both a finite target size and variable aperture size of the detector. The mathematical logic underpinning these simulations are shown in section 3, when considering isotropic point sources and point detectors.

As a result of these simulations, optimization procedures can be applied to determine the optimal target diameter and aperture size. These values could be applied to further investigations into experimental Compton scattering; results with the minimum possible cumulative uncertainty would be produced if the calculated optimal parameters are implemented into the experimental set-up.

## 2 Theory

### 2.1 Compton Scattering Equation

At its fundamental level, Compton scattering involves the scattering of photons by charged particles, here defined to be an electron at rest, where the transfer of both energy and momentum occur, such that the photon is scattered with reduced energy and momentum. The process incorporates both the conservation of momentum and relativistic mechanics; the energy transferred to the electron is comparable with its rest energy, and the process involves the scattering of massless photons.

$$E_\gamma = \frac{E_{\gamma_0}}{1 + \frac{E_{\gamma_0}}{m_e c^2}(1 - cos\theta)} \quad (1)$$

Considering the laboratory rest frame and applying both conservation processes for a photon scattered with angle θ, it can be shown in equation 1 that the scattered photon energy can be expressed as a function of the scattering angle, incident photon energy $E_{\gamma_0}$ and electron rest mass, $m_e$.

## 2.2 Experimental Compton Scattering

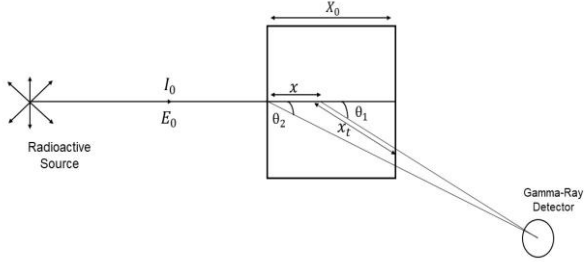

**Figure 1  Geometrical set-up of an isotropic point source and Compton scattering.** Photons are emitted with incident energy $E_0$ and intensity $I_0$ and scattered by a target of finite thickness $X_0$ to a gamma-ray detector, which may also be considered as a point source for simplicity. Adapted from Singh et al. 2006.

The photon scatters off the target into the detector, which itself consists of three main components: the scintillator, photomultiplier tubes (PMT) and the multi-channel analyzer (MCA). Visible light is formed in the NaI(TI) scintillator; this type specifically achieves a high light output and has an emission range compatible with the maximum efficiency of the bi-alkali photocathodes in the PMT. Light emitted from the scintillator enter the photocathode of the PMT, leading to a chain of photoelectrons being ejected; each electrode in the chain is biased more positively such that the emitted electrons are accelerated towards the next dynode, thus producing further secondary electrons. Electron numbers increase exponentially until they strike the anode at ground potential; electrons are collected and returned to ground through a resistor. Due to the monochromatic photons incident on the scintillator, a voltage pulse-height distribution is produced, with the height proportional to the energy deposited in the crystal via the absorbed gamma-ray.

The gamma ray spectrograph produces various features, the photopeak being the one of interest throughout this investigation, resulting from the complete photoelectric absorption of gamma rays into the scintillation crystal. Considering a Cs-137 source, other regions which arise, in order of increasing energy are: Ba-X ray peaks, backscatter peaks and a Compton edge. X-ray peaks occur within the Cesium source itself, as a result of internal transitions. A probable decay process is for the nuclei to transfer excitation energy to the K-shell electron, ejecting an electron from the atom; the vacancy in the K-shell thus results in X-ray emission [3]. Backscatter peaks are generally broad features and due to Compton scattering with the shielding, thus have a large scattering angle. Lastly, the Compton edge results from Compton scattering in the scintillator or detector. Thus, computational techniques are employed using topographic prominence to isolate the photopeak from each of these features.

## 2.3 Scattering Target

The geometric set-up of singly scattered photons with scattering angles $\theta_1$, originating from a target of thickness $X_0$ can be shown in figure 1 [4]. This forms the basis for both the simplified theoretical models discussed in 3 and the recommended adaptions to the investigation in 5. The theoretical number of photons at the detector from this set-up is expressed via equation 2 [5], using the incident photon flux $\phi$ and electron density $n_e$ in the target. Furthermore, the attenuation of incident photons in the target prior to scattering and scattered photons emerging from the target are expressed by $e^{-\mu_i x}$ and $e^{-\mu_t x_t}$, respectively.

The differential cross-section of photons scattered from a single free electron is introduced as a term in equation 2; this can be calculated via the Klein-Nashina equation for scattering angle θ, as shown in equation 3 [6]. Note that $P(E_\gamma, \theta)$ can be defined as the ratio of photon energies before and after the collision. At low frequencies, so therefore $E_\gamma \ll m_e c^2$, the value of $P(E_\gamma, \theta)$ tends to 1. This reduces to the classical expression and yields Thomson scattering, consistent with discussion in 1.

$$n_\gamma = \frac{\phi}{X_0} \int_0^{X_0} n_e \, e^{-\mu_i x} \frac{d\sigma}{d\Omega} \, e^{-\mu_t x_t} \, d\Omega dx \qquad (2)$$

$$\frac{d\sigma}{d\Omega} = 0.5\alpha^2 r_c^2 P(E_\gamma, \theta)^2 [P(E_\gamma, \theta) + P(E_\gamma, \theta)^{-1} - sin^2\theta] \quad (3)$$

## 3 Mathematics

### 3.1 Modelling Variable Target Size

Following from the theoretical description in section 2.3, an alternative two-dimensional scenario was considered such that the target can be modelled as a circle of radius $r$ with a point source and detector. This idea is represented in figure 2 [7], where two of the extreme angles of approach, their paths intersecting the target as shown to the top and bottom of the target, producing angles $\varphi_1$ and $\varphi_2$, respectively. This approach can be applied to the alternate scenario where the other two extremum points are instead placed on the right and left limits of the target, which form angles which are hereby denoted as angles $\varphi_3$ and $\varphi_4$, respectively.

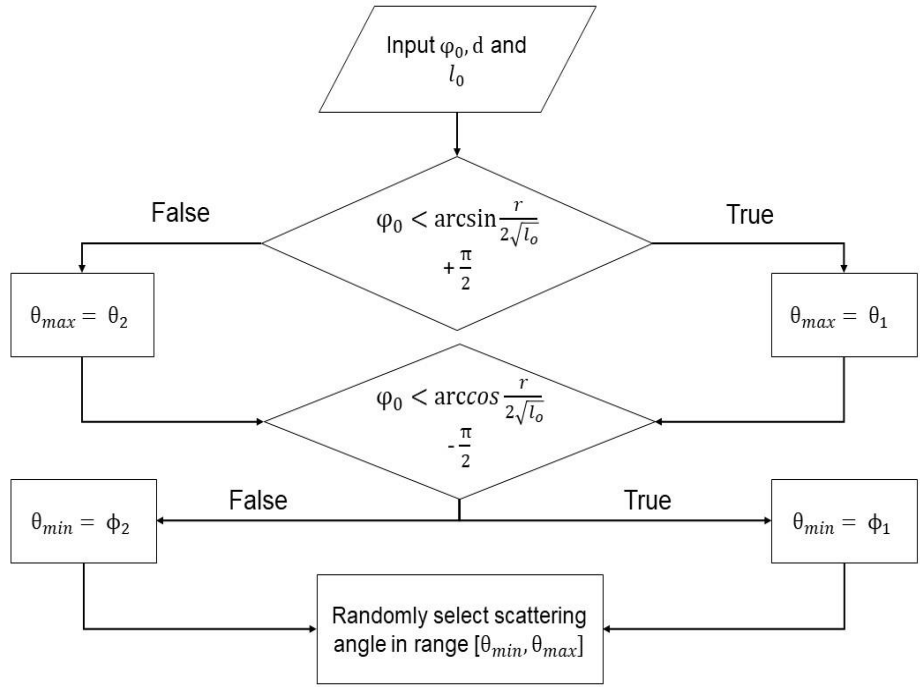

**Figure 3: The algorithm within the Monte Carlo simulation which calculates the range of scattering angles for a fixed experimental set-up.** Required inputs are the experimental angle, experimental path length, target radius and aperture width. This is numerous times, varying the latter two parameters simultaneously. The range of scattering angles is dependent on the value of the experimental angle with respect to the target radius and experimental path length, as derived from the mathematical models in 3. For each Compton scattering event, an angle is selected randomly from this outputted range.

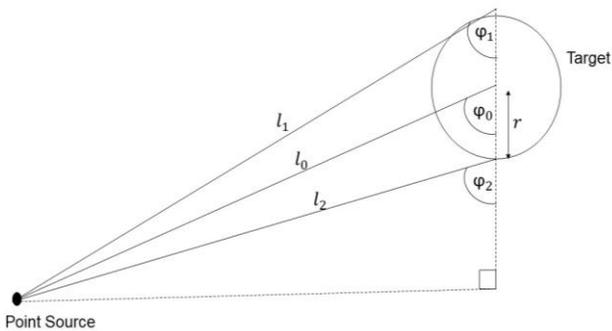

**Figure 2 Model of a point source and variable target size, underpinning the logic of incorporating finite target size into the Monte Carlo simulations**. The experimental mean angle is that subtended between the experimental mean path length, $l_0$ and the vertical line through the center of the target. Angles of this nature at the extreme limits of the targets $\varphi_i$ have accompanying extreme path lengths $l_i$. A similar application can be made to a second scenario where the extreme path lengths are at the left and right of the target from the center.

The scattering angle can hence trivially be defined as $2\pi - 2\varphi_i$; if all four extreme path lengths are considered, the maximum and minimum scattering angles, θ and ϕ, are summarized for each scenario in equation 4 and 5, respectively [8].

$$\theta_{1,2} = 2(\pi - \varphi_{1,3}) \quad (4)$$

$$\phi_{1,2} = 2(\pi - \varphi_{2,4}) \quad (5)$$

### 3.2 Modelling Variable Aperture Size

The model discussed in figure 2 can be extended to model for a variable aperture size of the detector. The greater the diameter of the aperture, there is a greater range of scattered angles available for detection, as opposed to an infinitely thin aperture producing a photopeak of the form of a Dirac delta function. Thus, there exist maximum and minimum scattering angles α allowed into the aperture of width $d$ as described in equation 6 [9].

$$\alpha_{max,min} = \varphi_0 \pm \arctan\frac{d}{2l_0} \quad (6)$$

## 4 Monte Carlo Simulations

Monte Carlo simulations provide a numerical solution based on random statistical trials, although as a result of its stochastic nature, a large number of events are required to achieve the desired statistical accuracy. Via this technique, simulations were developed to implement the variable target and aperture size; the relevant mathematical models and equations are expressed in 3.

The results of these simulations can be interpreted as the experimental measurements in the MCA; parameters to mimic an experimental set-up such as the resolution, bit-depth and gain of the experimental components are included. The simulated voltage-pulse distribution was then calibrated against known experimental evidence of Cs-137 scattering to produce a correct energy distribution. Thus, the simulation assumes a Cesium source.

Furthermore, errors arising from the experimental components were incorporated into the simulation. It can be noted that even gamma-rays of the same exact energy will not produce the exact same number of visible photons in the scintillator, thus different number of photoelectrons from the photocathodes in the PMT for each interaction. Ultimately, different numbers of secondary electrons produced at each dynode for each incident electron gives a distribution of produced voltage pulses and therefore, energies measured. A Gaussian distribution is obtained even for single energy gamma-rays, where a delta-function is expected, so noise was simulated around the photopeak to account for this effect.

The fixed experimental path lengths and angle selected for this investigation were 0.3m and 30°, respectively. From these input parameters, the range of scattering angles can be calculated, as demonstrated from the flow-diagram in figure 3. For each of the scattering events, the scattering angle is selected randomly from the outputted range.

## 5 Results and Discussion

From data observations, it was noted that for results with smaller values of the full-width half maximum (FWHM) of their simulated spectrographic photopeak, there were generally less detected events, as is logically expected. This in itself introduces another source of uncertainty: when applied to experiment, fewer detected events increases the difficulty of distinguishing the photopeak from the background count, or to isolate the peak from other spectral features in the spectrograph, as mentioned in 2.2. This factor was thus assigned an uncertainty $\pm 1/\sqrt{N}$, where $N$ is the total number of counts at the detector. A cumulative error is hence introduced such that it can incorporate both sources of error: FWHM of the photopeak and number of counts, which have an inverse relation to each other. This idea is demonstrated in figure 4, where the number of counts is expressed as either a function of the aperture width or target diameter It can be shown that for both variables, there exists saturations sizes at which there is no further change to the number of detected counts, indicated by the function tending to a constant value.

Both parameters can be varied simultaneously in an optimization procedure, in order to obtain the optimal set which produce the lowest possible cumulative uncertainty. Performing this procedure multiple times, the optimal values were computed to be 0.0215 ± 7 x $10^{-4}$ m and 5.35 x $10^{-3}$ ± 2.34 ± $10^{-5}$ m for target diameter and aperture width, respectively, yielding an uncertainty of 2.35 keV. These values are represented graphically in figure 5 for each parameter, where the minimum, thus optimum value is noted.

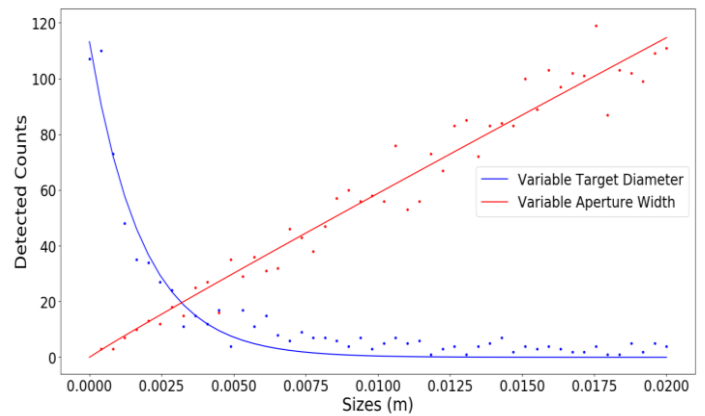

**Figure 4 Results from the Monte Carlo simulations when either target diameter or aperture width is varied, as indicated in the legend**. The other parameter remains fixed. Fits are produced for the collected data points; the count number decreases exponentially with increasing target diameter but increases approximately linearly with increasing aperture width until a discontinuity is reached, beyond which the number of counts becomes constant.

The uncertainties on the optimal parameters are in fact lower bounds, as there are experimental factors which have not been accounted for within the developed Monte Carlo simulation. Thus, there is likely to be discrepancies between these calculations and results obtained experimentally.

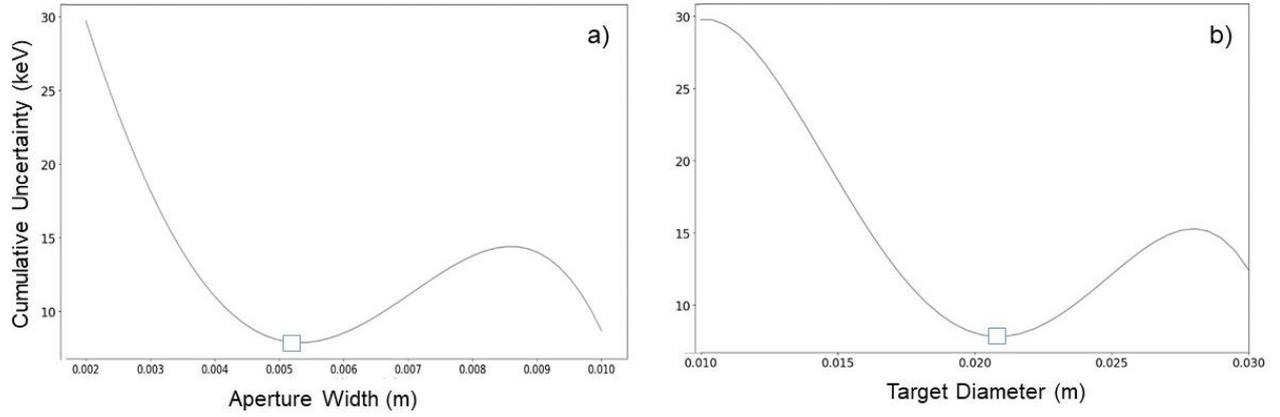

**Figure 5  The interpolated variation in the cumulative uncertainty with various parameter sizes with distinct minimum values**. These are the results of the Monte Carlo simulations, with a fixed experimental set-up of an experimental path length and angle of 0.3m and 30 °, respectively. The subplots are a) varying aperture width with the target diameter fixed at its calculated optimum value and b) varying target diameter for its optimized width. The minimum values can thus be shown to be 5.35 x $10^{-3}$ m and 0.0215 m for a) and b), respectively.

Particularly, the experimentally measured number of counts at the detector for each experimental mean angle may be significantly less than the simulated results. The simulation assumes infinitely thin targets, which is the optimal scenario, but impossible to implement physically. This simulation is thus restricted to 2D, but further investigations can be made to extend the simulation to 3D; one of these methods by considering the thickness of the scintillation crystal, thus providing more physically accurate results. Hence, the results would become more applicable to future experimental Compton scattering investigations, where a minimum cumulative uncertainty in the set-up would be required.

$$\varepsilon_i = 1 - e^{-\mu_{tot} E t} \quad (7)$$

A technique to achieve this is to account for the response function of the NaI(TI) scintillator. The intrinsic crystal efficiency $\varepsilon_i$ for a crystal thickness $t$ and photopeak energy $E$, as expressed in equation 7 [10], is simply assumed to be 1. Thus, the term accounting for the attenuation coefficient of the crystal is neglected. The number of detected counts can then be adjusted by a factor of $\varepsilon_i$. The results should follow a relation closer to that of equation 2; extensions can hence be made that account for the both stated attenuation factors for a photon entering and leaving the target, as well as incorporating the Klein-Nashina equation described in equation 3.

Additionally, only singly scattered photon events are considered. Multiply scattered photon events can occur, these increase with the mean path length through the target, thus the target thickness. These simulations could be extended to 3D as a further investigation, such to model the saturation depth at different incident energies; this is 20mm for 0.662 keV in a Copper target [11]. The signal to noise ratio can thus be defined as the number of singly scattered and multiply scattered events. With increasing target thickness, the signal-to-noise ratio decreases; thus, target thickness could be incorporated and optimized amid other various constraints.

## 6 Conclusion

The necessary mathematical models for variable target size and aperture width in experimental Compton scattering are derived and presented, which were subsequently incorporated into a computational Monte Carlo simulation. The optimal values of the parameters which produce the minimum possible uncertainty value in the photopeak energy of 2.35 keV were found to be 0.0215 ± 7 x $10^{-4}$ m and 5.35 x $10^{-3}$ ± 2.34 ± $10^{-5}$ m for target diameter and aperture width, respectively. Recommendations for further investigations by generalizing the simulation to three dimensions are made, by incorporating the thickness of the scintillation crystal and accounting for the attenuation factors.

**Acknowledgments**

I would like to thank the Dr Steve Kolthammer, who provided much intellectual discussion on this topic and my lab partner, Nathalie Podder, whose work was vital for creating the simulations.

**References**

[1] A.H Compton, *A Quantum Theory of the Scattering of X-rays by Light Elements*, Physics Review Volume 21, Issue 483, (1923)

[2] W. Bothe, H. Geiger: *Über das Wesen des Comptoneffekts; ein experimenteller Beitrag zur Theorie der Strahlung*. Zeitschrift für Physik **32**, 639–663, 1925

[3] J.E Parks, *The Compton Effect-Compton Scattering and Gamma Ray Spectroscopy*, University of Tennesee, Tenesee, 2015, available from: http://www.phys.utk.edu/labs/modphys/Compton%20Scattering%20Experiment.pdf [accessed 22/2/21]

[4] Singh et al., *Energy and Intensity Distribution of Multiple Compton Scattering*, Physical Review A, Issue 74, 2006

[5] Gallardo S, Ródenas J, Verdú G. *Monte Carlo Simulation of the Compton Scattering Technique applied to Characterize Diagnostic X-ray Spectra*. Med Phys. 2004, pg 2082-90. doi: 10.1118/1.1759827. PMID: 15305461

[6] Kalyvas, N., Valais, I., David, S. *et al. Studying the Energy Dependence of Intrinsic Conversion Efficiency of Single Crystal Scintillators under X-ray Excitation*. Opt.Spectrosc. **116,** 743–747(2014). https://doi.org/10.1134/S0030400X14050117

[7] A.S Felice, *Third Year Laboratory Book,* Blackett Laboratory, Imperial College London, 2021, pg 16

[8] A.S Felice, *Third Year Laboratory Book*, Blackett Laboratory, Imperial College London, 2021 pg. 25

[9] A.S Felice, Third Year Laboratory Book, Blackett Laboratory, Imperial College London, 2021, pg.32

[10] L. Jarczyk, H. Knoepfel, J. Lang, R. Müller, W. Wölfli, Photopeak Efficiency and Response Function of *Various NaI(Tl) and CsI(Tl) Crystals in the Energy Range up to 11 MeV*, Nuclear Instruments and Methods ,Volume 17, Issue 3,1962, Pages 310-320, ISSN 0029-554X, https://doi.org/10.1016/0029-554X(62)90009-5.

[11] W. Mannhart, H. Vonach, *Gamma-ray absorption coefficients for NaI(Tl),* Nuclear Instruments and Methods,Volume 134, Issue 2,1976, Pages 347-351, ISSN 0029-554X, https://doi.org/10.1016/0029-554X(76)90291-3.